\documentclass[aps,pre,amsmath,amssymb,twocolumn,showpacs,byrevtex,superscriptad
dress]{revtex4}
\def\figwidth{0.9\linewidth}
\usepackage[final]{graphicx}
\usepackage{epstopdf} 
\begin{document}
\title{On Static and Dynamic Heterogeneities in Water} 
\def\roma{\affiliation {
Dipartimento di Fisica and INFM Udr and CRS-SOFT: Complex Dynamics in Structured Systems, Universit\`a 
di Roma ``La Sapienza'', Piazzale Aldo Moro 2, I-00185, Roma, 
Italy}}
\author{Emilia La Nave}\roma
\author{Francesco Sciortino}\roma

\begin{abstract}
We analyze differences in dynamics and in properties of the sampled potential energy landscape between different equilibrium trajectories, for a system of rigid water molecules interacting with a two body potential.  On entering in the supercooled region, differences between different realizations enhance and survive even when particles have diffused several time their average distance.  We observe a strong correlation between the mean square displacement of the individual trajectories and the average energy of the sampled landscape.
\end{abstract}

\maketitle

\section{Introduction}
\label{intro}
    
The  Potential Energy Landscape (PEL)  \cite{wales,pablonature} is the surface 
generated by the potential energy of a system. In the case of system composed by $N$ rigid  molecules, it is an highly complex surface defined in a $6N$ dimensional space.   In recent years, numerical studies  \cite{skt99,scala00,sastry01,starr01,speedyjpc,otplungo,wales2,lacks,heuernew},  boosted by the increased numerical resources, have attempted to quantify the statistical
 properties of this surface (for example quantifying the distribution in energy of specific points of the surface, like local minima and saddles) in the attempt to develop a thermodynamic description fully  based on PEL properties. This line of thinking, 
pioneered by Frank Stillinger and his coworkers \cite{stillinger_pes},
  has been very fruitful in the study of supercooled liquids,
 both in equilibrium  ~\cite{scala00,sastry01,voivod01,otplungo} and in 
 out-of equilibrium conditions ~\cite{st,crisanti,parrocchia,scottjcp}.
   Stillinger's formalism  builds on the idea that the PEL 
 surface can be partitioned into disjoint basins. A basin is unambiguously defined as the 
 set of points in configuration space connected to the same local minimum  --- named    
inherent structure (IS) --- via a steepest-descendant  trajectory.  
In this respect, the  PEL's statistical properties entering in the evaluation of the partition 
function are the number, shape and depth  of the PEL basins.  For its conceptual simplicity 
and its strict connection with  numerical implementation, the PEL formalism has become one of the modern tools to interpret and analyze simulation data.

Water, due to its intrinsic interest as liquid of life, 
has been extensively studied in computer simulations. Models have
 been developed which are able to reproduce qualitatively the thermodynamic
 and dynamic anomalies of this liquid \cite{spce,t5p,st2}.
Indeed, water is characterized by a line of isobaric density maxima and compressibility  minima. Its specific heat $C_{p}$ increases on cooling~\cite{angellreview,pablobook}.   
 Recent studies have attempted to connect the thermodynamic  anomalies to
 the presence of  a line of first order transition between two liquid structures, 
 eventually ending into  a  second order critical  point ~\cite{poole,ponyatonski}.  
Dynamics in water also shows anomalies.  The diffusion coefficient decreases as pressure
 is increased up to a maximum value (approximately 400 MPa\cite{lang} ) at 
room temperature.  For greater  pressures, water dynamics becomes progressively slower, 
as expected for simple liquids.

 Water simulations have been among the first to be scrutinized using the methods developed by Stillinger\cite{stillingerwater,ohmine1,ohmine2,nature5neigh,scala00,starr01,jcpmio} in  the attempt to clarify differences in  local  structures and  relations between structural properties and dynamics.  More recently, extensive studies of the  PEL properties have been published, providing a detailed description of the landscape properties of several models of water\cite{starr01,debenewater,genebirt,nicolas,spceprl}.
Within the PEL formalism, attempts have also been presented in the direction of connecting thermodynamic properties (like the number of explored 
minima or the number of diffusive directions in the PEL) to dynamics
 \cite{tom,inm,angelani,scala00,otplungo,silicaprl,heuer,cavagna,doliwa,schroder,harrowell}.
The outcome of these works, and related studies on different models of glass forming liquids, suggests a strong connection between dynamics 
and landscape properties. Still, the interpretation schemes require fitting parameters
whose physical interpretation in often unclear. As an example, it
 has been shown  that $ln(D)$ is consistent with a  $1/(TS_{conf})$  dependence \cite{scala00,sastry01,otplungo,voivod01,cristiano} (being $S_{conf}$ the configurational entropy) but no understanding  of the proportionality coefficient in term of PEL properties has been reported \cite{ruoccojcp}.   

In this article  we report analysis of distinct equilibrium trajectories of
a system of 216 water molecules interacting via a two body potential.
 The novel aspect is  the possibility of comparing a large
 number of independent trajectories (more than 100 for each state point)
 for a large number of state points. Each of these (previously equilibrated)
 independent trajectories last more than 20 ns, at which time --- even at the 
slowest studied temperature --- the average displacement of the molecule is
 longer than three nearest neighbors.
  The analysis of this set of data shows that, while individually each trajectory
 appear to be to a very good approximation a diffusive trajectory 
(i.e. with a mean square displacement 
which increases linearly with time) different realizations are
 characterized by apparent values of the diffusion coefficient which differ by more than one order of magnitude, 
a clear evidence of dynamic heterogeneities. We also show that 
differences in dynamics are strongly coupled with the location
 of the system on the PEL, providing evidence that dynamic heterogeneities
 are related to inhomogeneities in the
local basin energies.   For completeness, we briefly discuss
 the distribution of IS energies for this model  and the temperature and density ($\rho)$ dependence of the
diffusion coefficient, presenting new data with significantly improved statistics.

\section{Simulation Details}
\label{simulation}

We analyze molecular dynamics trajectories of a 
system composed of  216 rigid water molecules in the (NVE) ensemble. 
Molecules interact via the widely studied two-body  simple point charge 
extended (SPC/E) model \cite{spce}. The integration step is 1 $ft$ second, 
and long range interactions are taken in account using the reaction  field method.
 Dynamics and thermodynamics properties for the SPC/E model have been extensively   studied in the past \cite{pablo} in a wide range of temperatures and densities.  Analysis of the statistical properties of the potential energy landscape have also  been characterized \cite{debenewater,scala00,starr01,nicolas,genebirt,spceprl}. Here we analyze a large set of data, based on 5 different temperatures
 in supercooled states and nine different densities, from 0.9 to 1.4 g/cm$^3$. At each state  point, we analyze equilibrated  trajectories of more than 100 independent realizations, to
 both reduce significantly the numerical error and to estimate the self-averaging
 properties of this model for different $T$ and $\rho$. 
 Each trajectory covers a time interval of about 20 ns.   Potential energy landscape properties have been based on the analysis of the 
 inherent structures, which has been calculated 
   using a standard conjugate gradient minimization algorithm with $10^{-15}$ tolerance \cite{numericalrec}. 

\section{IS Energies: the random energy model}
\label{is-prop}
A quantification of the statistical properties of the potential energy landscape,
 i.e. the distribution of basin's depth and shapes,
 for the SPC/E model of water has been recently reported\cite{spceprl}.
 It has been shown that  the number $\Omega (e_{IS})de_{IS}$ of distinct basins
 of energy depth  $e_{IS}$  between $e_{IS}$ and $e_{IS} + de_{IS}$  follows 
a Gaussian distribution,

\begin{equation}
\Omega(e_{IS}) de_{IS}=e^{\alpha N}
\frac
{e^{-(e_{IS}-E_o)^2/2\sigma^2}}
{\sqrt{2 \pi \sigma^2}}
de_{IS},
\label{eq:Omega}
\end{equation}

\noindent
where $e^{\alpha N}$ is the total number of distinct
basins of the PEL for the system of $N$ molecules,
$E_0$ is the energy scale of the distribution and $\sigma^2$ is the variance. 
The coefficients $\sigma$, $E_o$ and $\alpha$ depend only on the system density. 
The corresponding probability distribution $P(e_{IS},T)$  of  sampling an $IS$ of depth $e_{IS}$  in equilibrium at temperature $T$ is given by
\begin{equation}
P(e_{IS},T)=\frac{   \Omega(e_{IS})   e^{
- \beta   [e_{IS} +f_{vib}(T,e_{IS})] }}
{\int P(e_{IS},T)de_{IS} },
\label{eq:peis}
\end{equation}
\noindent
where $\beta=1/kT$ and $f_{vib}(T,e_{IS})$ is the  basin  free energy \cite{skt99,sastry01}. 
  The hypothesis of a Gaussian form for $\Omega(e_{SI})$, together with the
assumption of an  $e_{IS}$ independence of the 
basin anharmonicities implies: (i)  that the $T$ dependence  of the average $IS$ energy  $<e_{IS}(T)>$, at constant volume, is linear  as a function of $T^{-1}$\cite{sastry,otppisa,eos}, that is:
\begin{equation}
<e_{IS}(T)>= A+\frac{B}{T},
\label{eq:ab}
\end{equation}
where the coefficients $A$ and $B$ depend only on $\rho$ and can
 be expressed in term of landscape properties\cite{eos},
 and (ii) that the probability distribution $P(e_{IS},T)$  is Gaussian with a $T$-independent variance\cite{sastry,otppisa,eos,sasai,rem,tomrem,ivan}.

\vskip 0.8cm
\begin{figure}[th]
\begin{center}
\includegraphics[width=\figwidth]{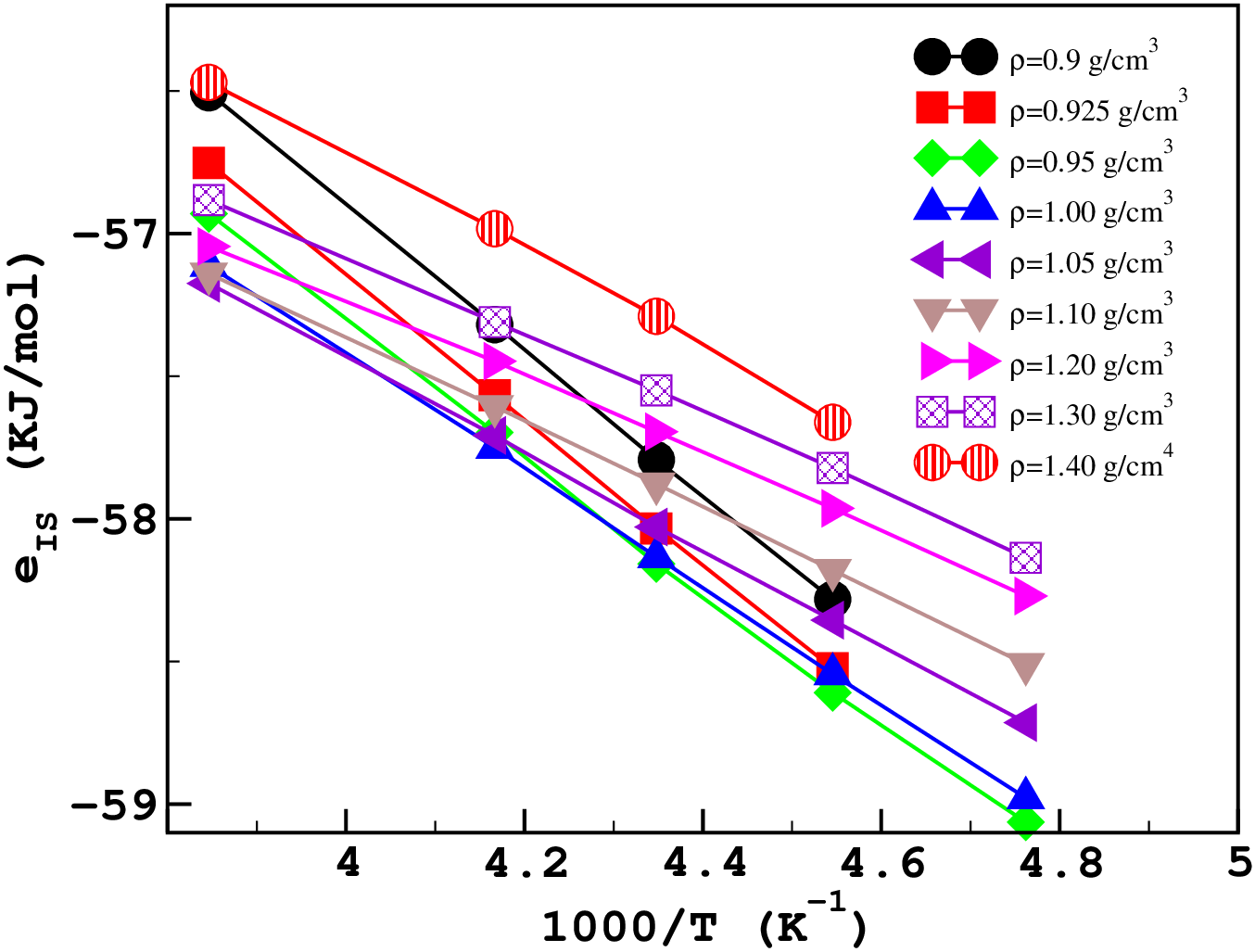}
\end{center}
\caption{Temperature  dependence of the average $<e_{IS}>(T)$ for all investigated densities.}
\label{fig:eisdit}
\end{figure}

\begin{figure}[h]
\includegraphics*[width=\figwidth]{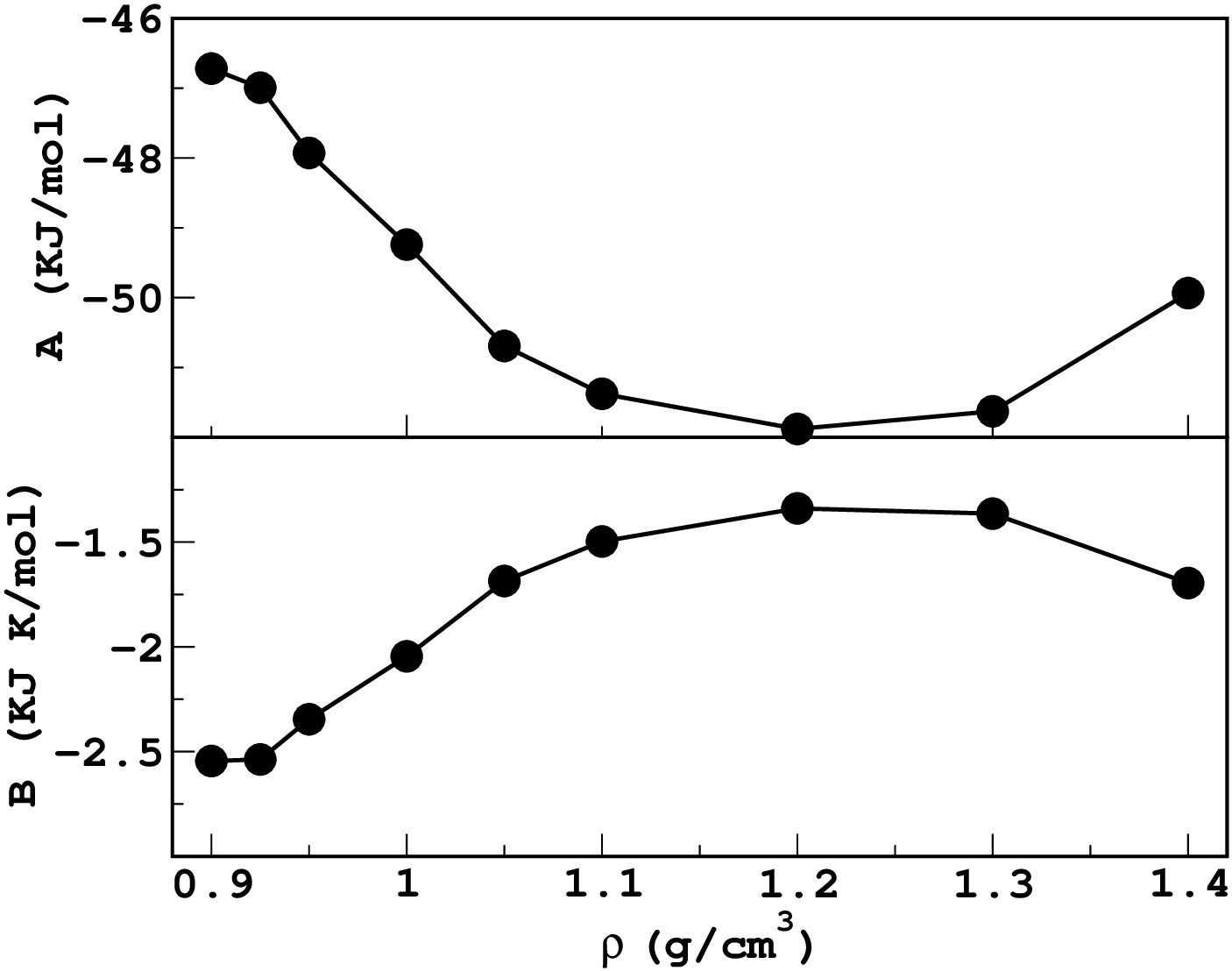}
\caption{Density dependence of the fitting parameters $A$ and $B$ in Eq.\ref{eq:ab}}
\label{fig:parametri}
\end{figure}

To provide evidence of this behavior  we report in Fig. \ref{fig:eisdit}  $<e_{IS}(T)>$ at all studied densities. In all cases, in the investigated $T$-range, the expected $T^{-1}$-dependence holds. 
Nevertheless, two breakdowns of Eq.\ref{eq:ab} are expected  outside the investigated $T$-region. At high $T$,
 due to the increasing importance of the anharmonic contribution to the basin free
 energy\cite{otppisa} and at low $T$ due to possible presence of non-Gaussian corrections to the  $e_{IS}$ probability distribution (Eq. \ref{eq:Omega}). The low $T$ region can not be numerically studied due to the huge increase of 
 the  relaxation times. 
Fig. ~\ref{fig:parametri} shows the $\rho$ dependence of  $A$ and $B$ (which have well defined landscape interpretations\cite{eos}).  As a first approximation (but see Ref.\cite{otppisa} for a more precise discussion), $A$ is related to  $E_0$ and  $B$ to $\sigma^2$.

\begin{figure}[h]
\centering
\includegraphics*[width=\figwidth]{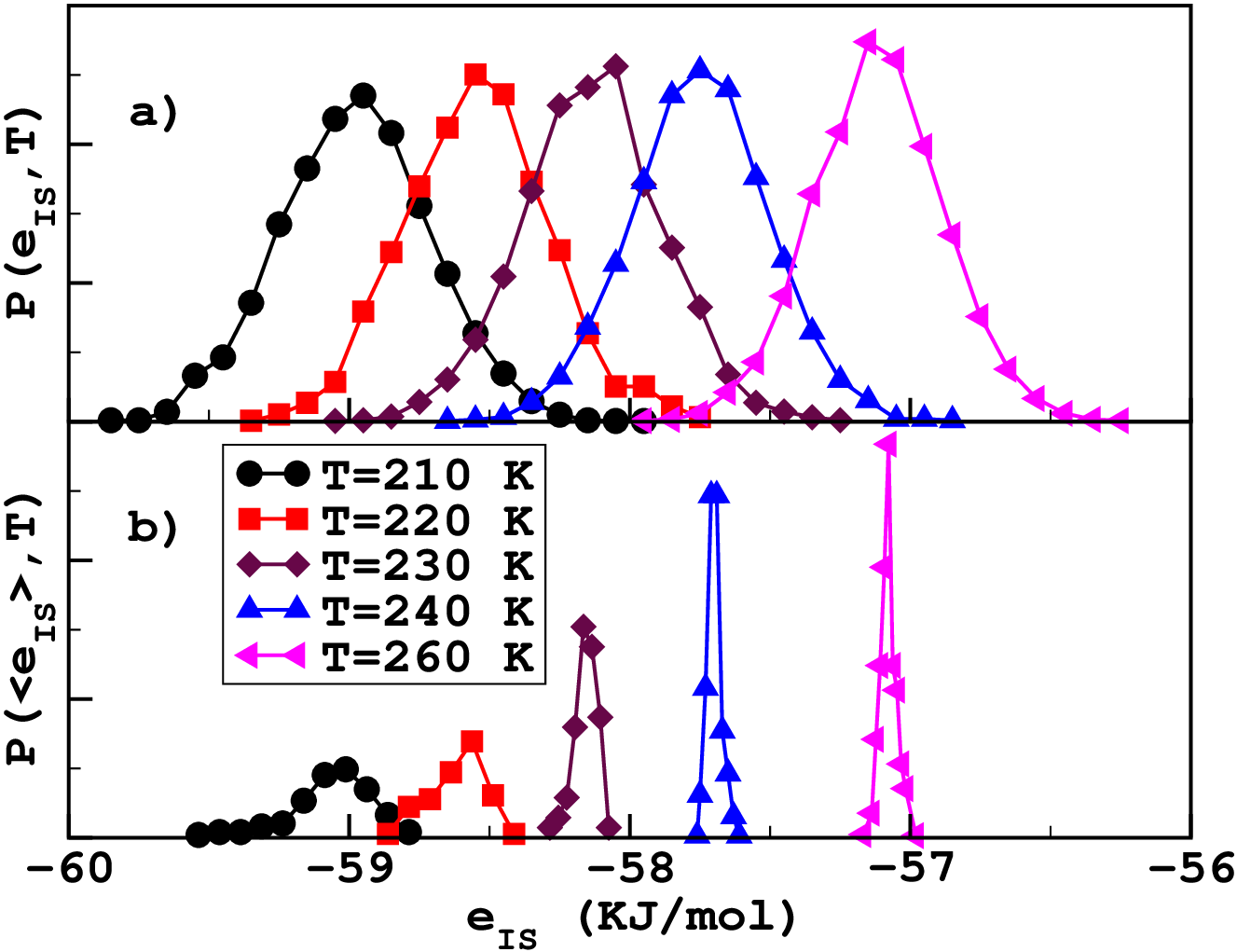}
\caption{a) Probability distribution of the $e_{IS}$ (per molecule)  at density $d=0.95$ g/cm$^3$, for different $T$ values. b) Probability distribution of $<e_{IS}>$ , where each average is calculated  over one distinct trajectory. Units are arbitrary.
Each $P(e_{IS},T)$ is calculated using
$5000$ points, since from each of the 100 independent trajectories we evaluated 
50 different inherent structures,  minimizing at intervals equally spaced in time. 
}
\label{fig:dist}
\end{figure}

\begin{figure}[h]
\centering
\includegraphics*[width=\figwidth]{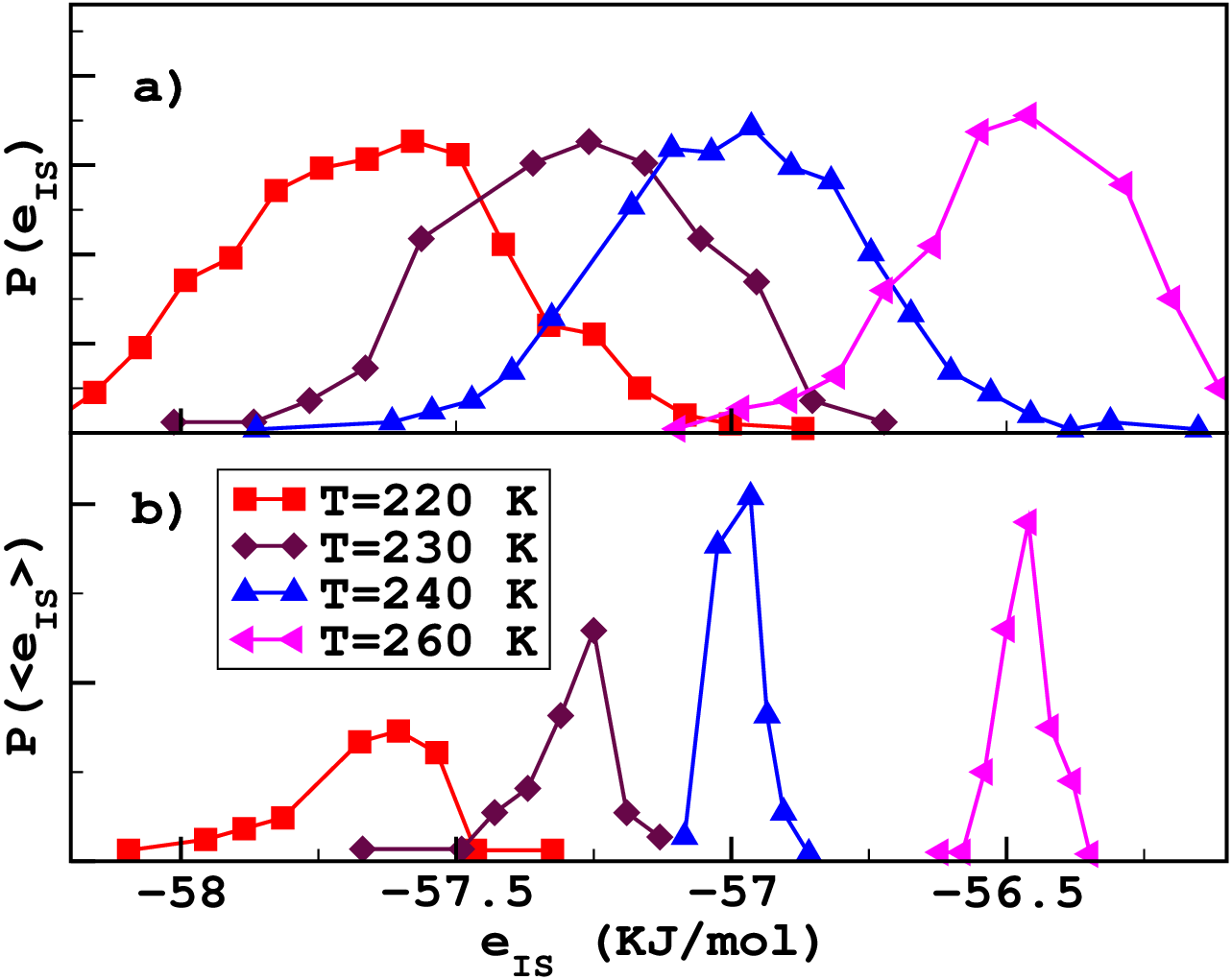}
\caption{Same as Fig.\protect\ref{fig:dist} for $d=1.40$ g/cm$^3$.}
\label{fig:dist2}
\end{figure}

The large available data set  allows us to evaluate $P(e_{IS},T)$ and test the second prediction  of the Gaussian hypothesis.  Fig.~\ref{fig:dist}-a and \ref{fig:dist2}-a
  show $P(e_{IS},T)$ respectively for  $\rho=0.95$ $g/cm^3$ and $\rho=1.00$ $g/cm^3$ . 
Figures show that  the shape of the distribution is, to a first 
approximation, Gaussian and that the variance of the distribution only weakly increases with  $T$, as can be inferred by the small change of the heigh of the distribution with $T$.

\begin{figure}[h]
\centering
\includegraphics*[width=\figwidth]{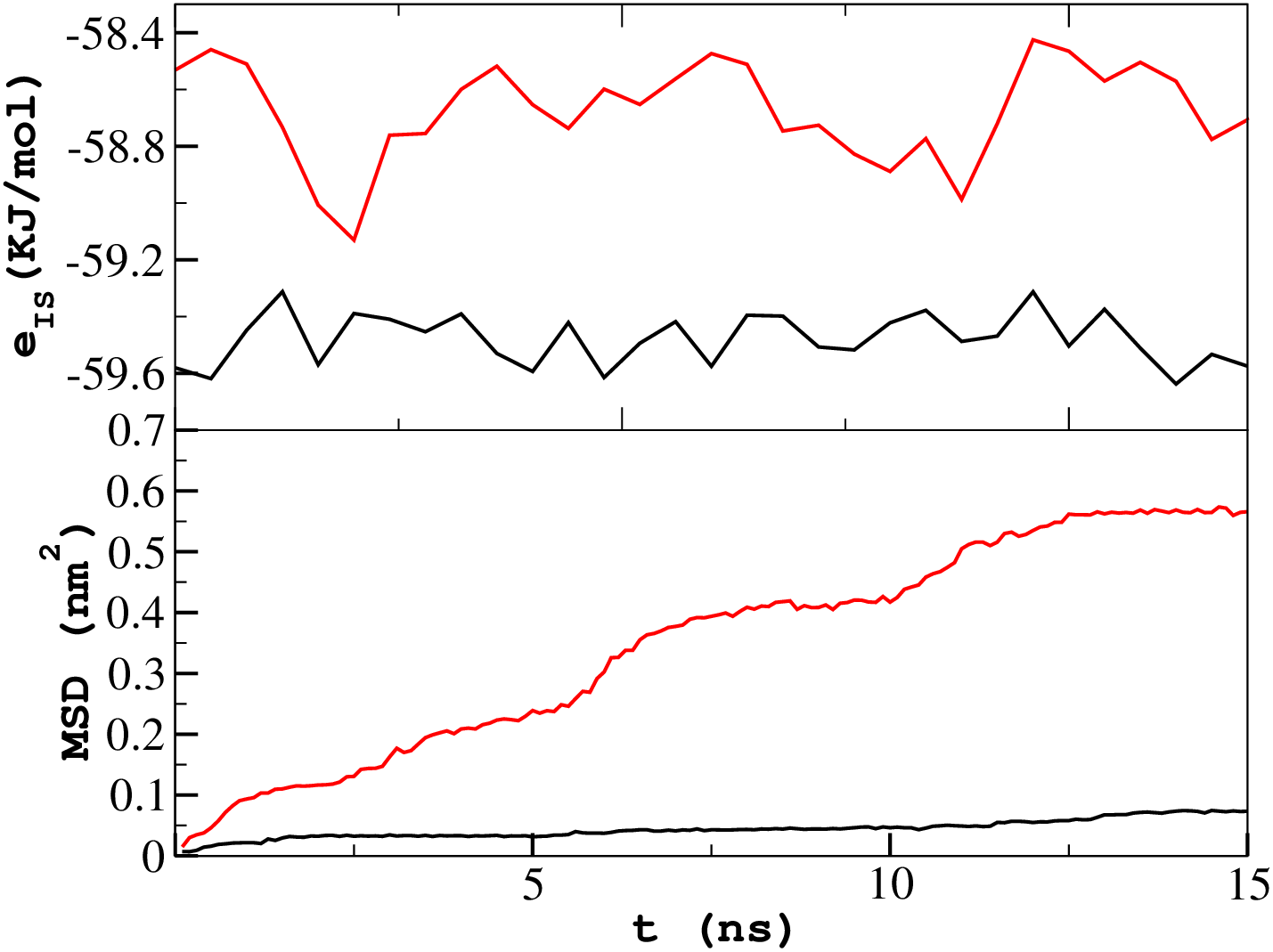}
\caption{ $e_{IS}$ and the mean square displacement  as a function of time for two different trajectory
at  $T=210  K$ and $\rho=0.95 $ g/cm$^3$.}
\label{fig:tvseis}
\end{figure}

The availability of 100 independent realizations allows us to study also the distribution
 of $P(<e_{IS}>_i, T )$, defined as the
distribution of the average depth sampled in each independent trajectory $i$ in the simulated 20 $ns$ time interval. In this case, of course,
 each distribution is evaluated only over 100 points.  If the length of each trajectory is sufficiently long, so that the system is able to sample all basins which are  statistically relevant at that temperature, 
the distribution should be  peaked  around the average of $P(e_{IS},T)$, with a small variance.
  Fig.~\ref{fig:dist}-b and Fig.~\ref{fig:dist2}-b
show that this is the case only at the higher $T$.  On cooling, the relaxation time of  the system increases and, within the time of
 the simulation, the system retains memory of the initial basin.  It is important to observe that the distribution becomes very asymmetric,
 developing a long tail at low basin energies. 
This may suggest a strong relation between 
depth of the explored basin and dynamics, which can not be attributed to differences in
 thermal energy, since all different realizations have the same kinetic energy.
 In particular, the asymmetry in the resulting distribution suggests  that configurations starting  from low energy basins do not have time to explore phase space sufficiently.  It is worth stressing, that similar results are 
observed at all investigated densities.  To support   this hypothesis we show in Fig.~\ref{fig:tvseis}  the time evolution of $e_{IS}$ and of the center of mass mean square displacement $MSD_i$  for two different trajectories at the same $T$  and $\rho$.  
More precisely, the mean square displacement of the trajectory $i$ is defined as
\begin{equation}
MSD_i= \frac{1}{N} \sum_{j=1}^N [\vec r^{CM}_j(t)-\vec r^{CM}_j(0)]^2, 
\end{equation}
\noindent where $\vec r^{CM}_j(t)$ is the location of the center of mass of molecule $j$ in the trajectory $i$  at time $t$. The averaged $MSD$
is defined as 
\begin{equation}
MSD=\frac{1}{{\cal N}_t} \sum_{i}^{{\cal N}_t} MSD_i,
\end{equation}
\noindent where 
${\cal N}_t$ is the number of independent trajectories.
The two  chosen trajectories
differ by the value of $e_{IS}$ at time $0$.
In the  studied $t$ range ($ns$),  both configurations sample a restricted interval of $e_{IS}$ values. The memory of the  energy of the starting basins is preserved during the 
simulation time. Comparing Fig.~\ref{fig:tvseis}-a and Fig.~\ref{fig:tvseis}-b we note that the
configuration with large $e_{IS}$ is characterized by a larger $MSD$.

\section{Dynamic Heterogeneities}

\begin{figure}
\includegraphics*[width=\figwidth]{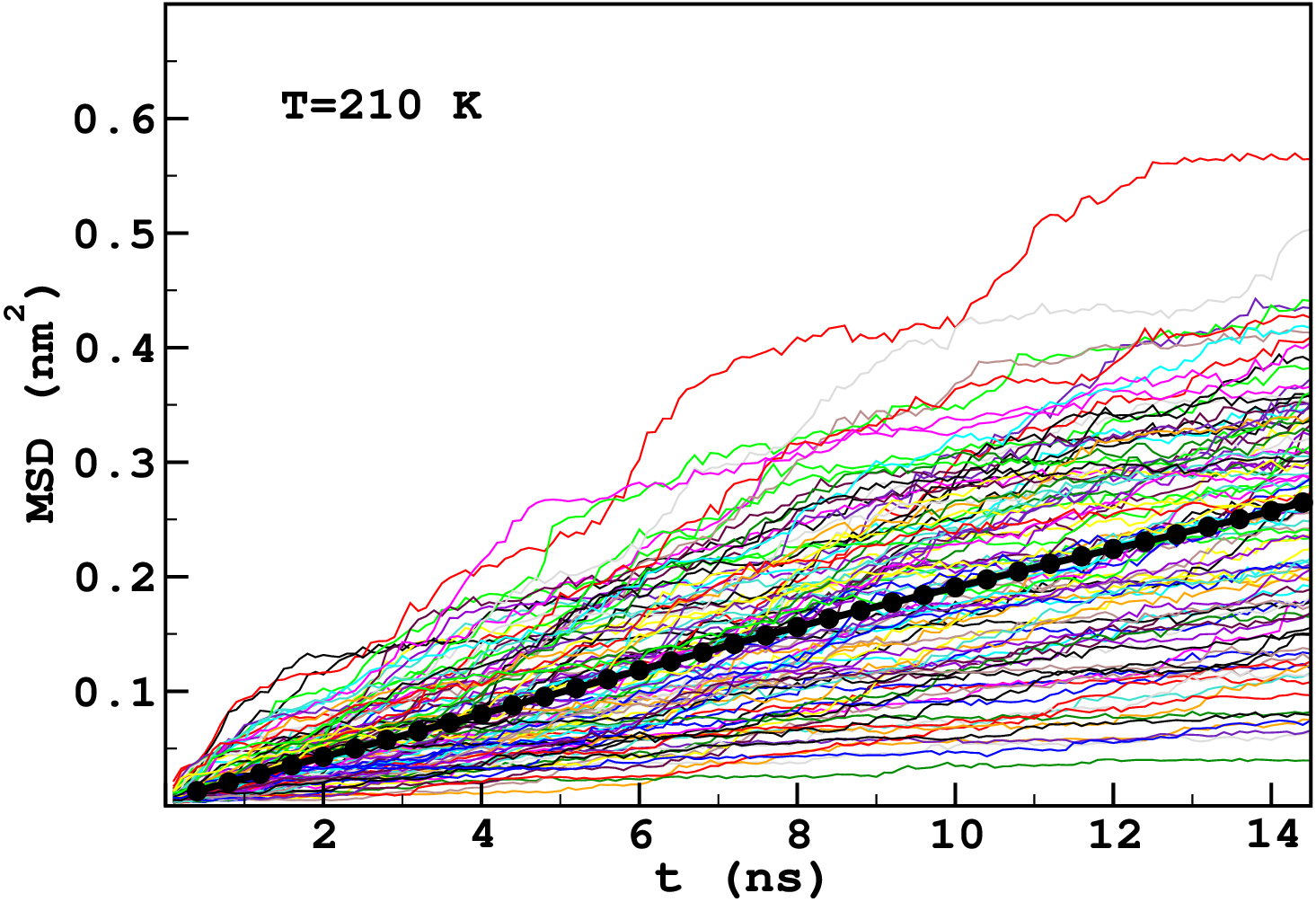}
\includegraphics*[width=\figwidth]{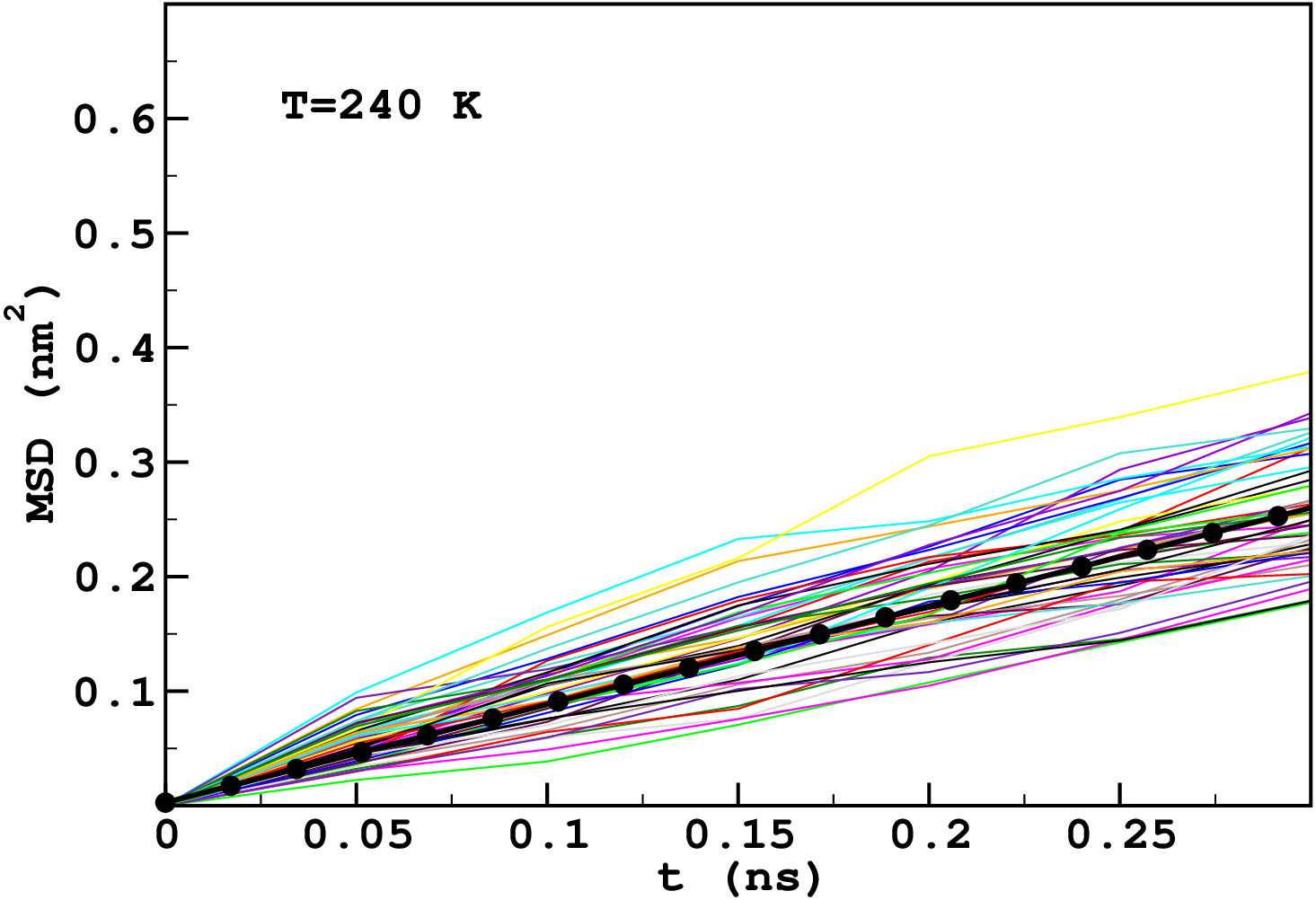}
\includegraphics*[width=\figwidth]{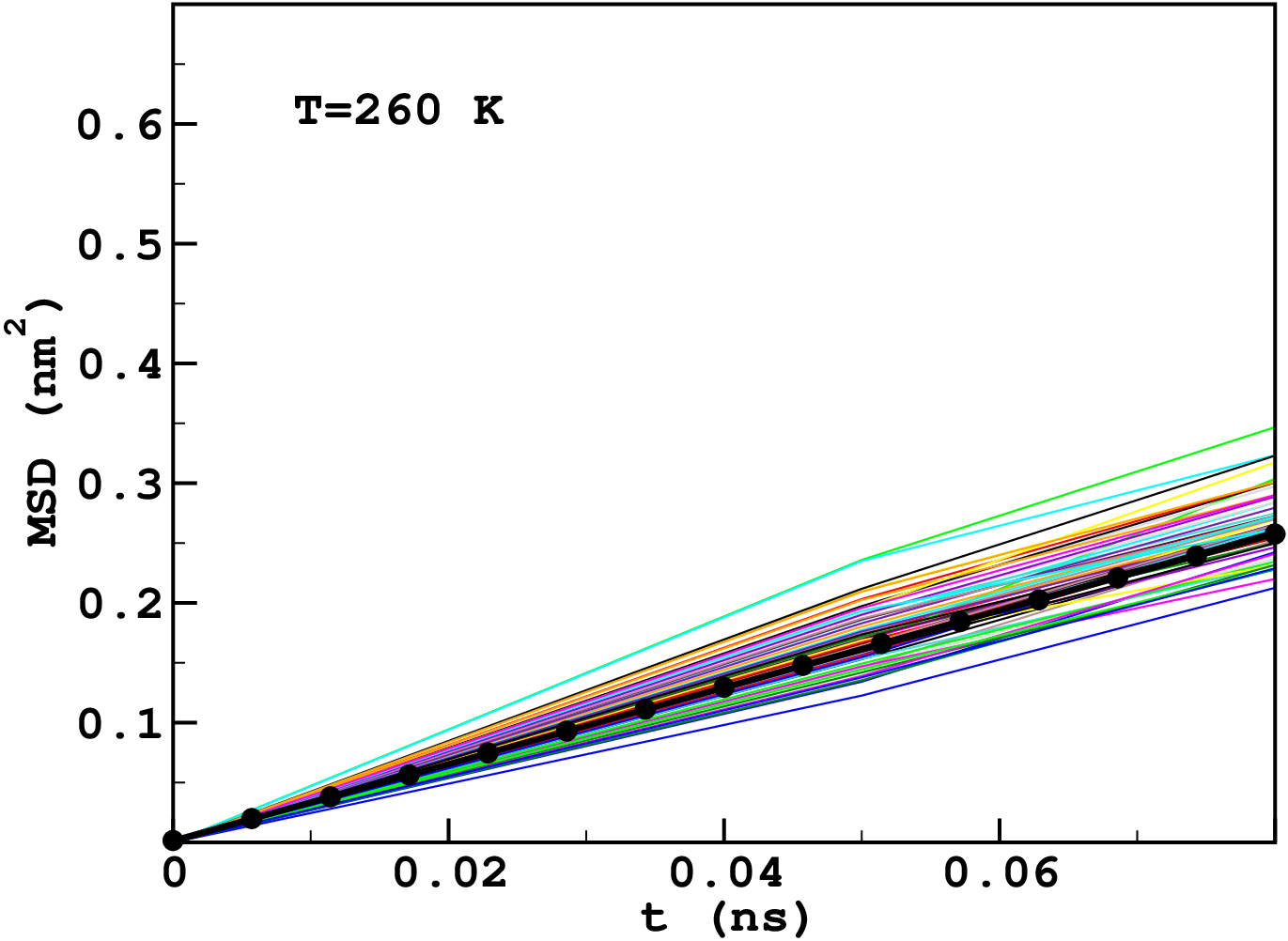}
\caption{Individual mean square displacement for 100 independent realizations at three different temperatures. Filled symbols indicate the $MSD$ averaged over the different realizations. Note that no average over starting time has been performed. }
\label{fig:newfig}
\end{figure}

We next focus the attention on dynamic heterogeneities via a study of the 
mean square displacement  of the individual trajectories\cite{tom,tom2}. 
For each state point $(T,\rho)$  we calculate  $MSD_i$  for each of the ${\cal N}_t$ trajectories. Fig.~\ref{fig:newfig}  shows  $MSD_i$  for $\rho=0.95$ g/cm$^3$ at three selected $T$, but similar results are obtained for all the other studied densities. In all cases, no averaging over different time origins is performed.  In the figure, the time axis  are chosen in such a way that the average $MSD$ at the maximum reported time coincides for all temperatures. A comparison of the spreading of the different realizations at fixed value of the average $MSD$, for example at the maximum reported time, reveals a clear increase  in the fluctuations of the different trajectories on lowering $T$. 

To better quantify fluctuations in dynamics   and the role of $T$, we calculate the variance of the mean square displacement of different trajectories. More precisely, we evaluate,  for each time $t$ 
\begin{equation}
\sigma_{_{MSD}}=\sqrt{\frac{\sum_i (MSD_i-MSD)^2}{{\cal N}_t-1}}.
\end{equation}

\begin{figure}
\includegraphics*[width=\figwidth]{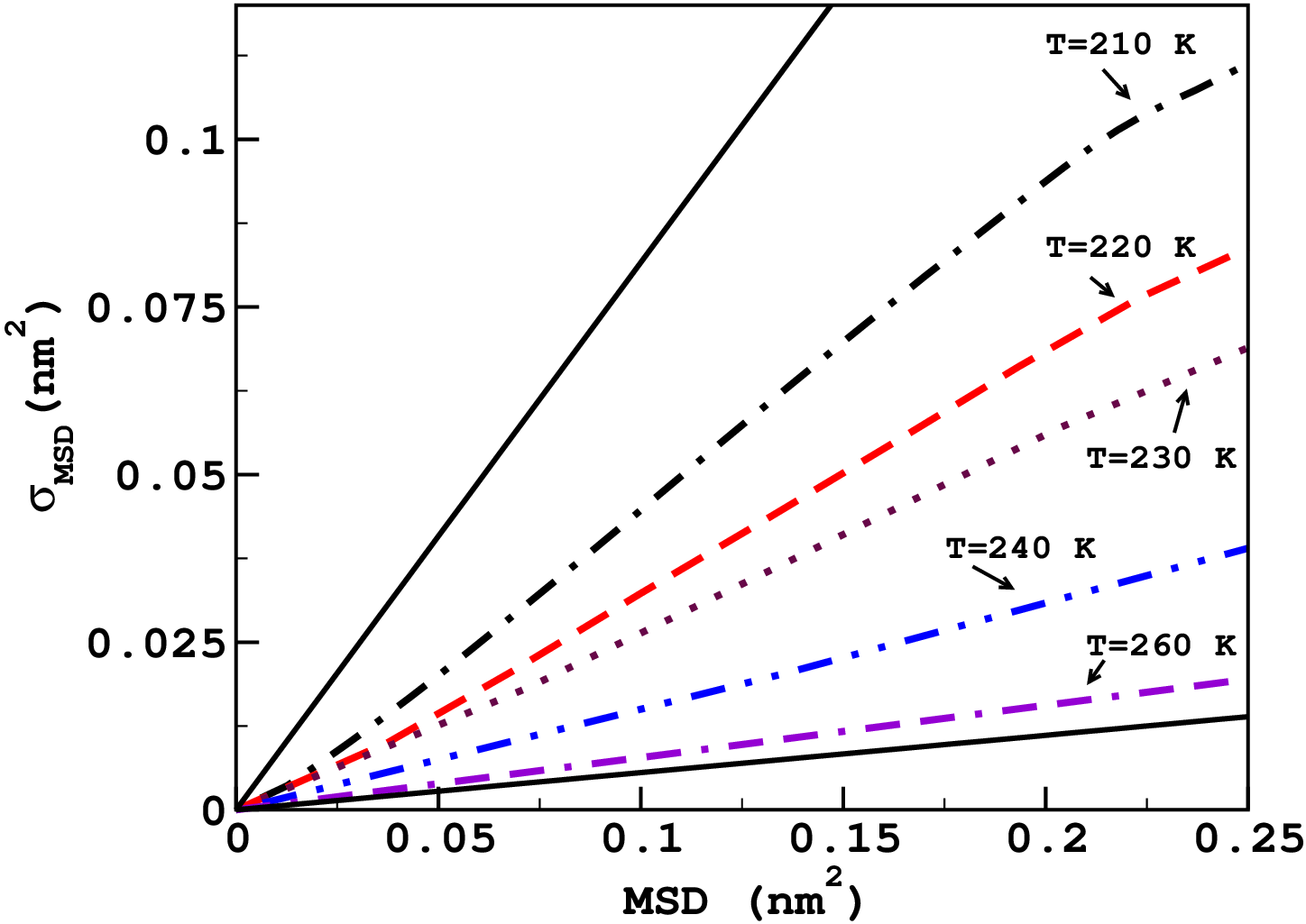}
\caption{Variance $\sigma_{_{MSD}}$ of the $MSD $ as a function of the average MSD for $\rho=$ 0.95 g/cm$^3$, for different $T$.  
The two full lines indicate the two extreme limits provided by Eqs.~\ref{eq:highT}-\ref{eq:lowT}.
Note that in water the nearest neighbor distance is
0.28 nm, corresponding to a square displacement of $\approx 0.09$ nm$^2$. }
\label{fig:spread}
\end{figure}

\begin{figure}
\includegraphics*[width=\figwidth]{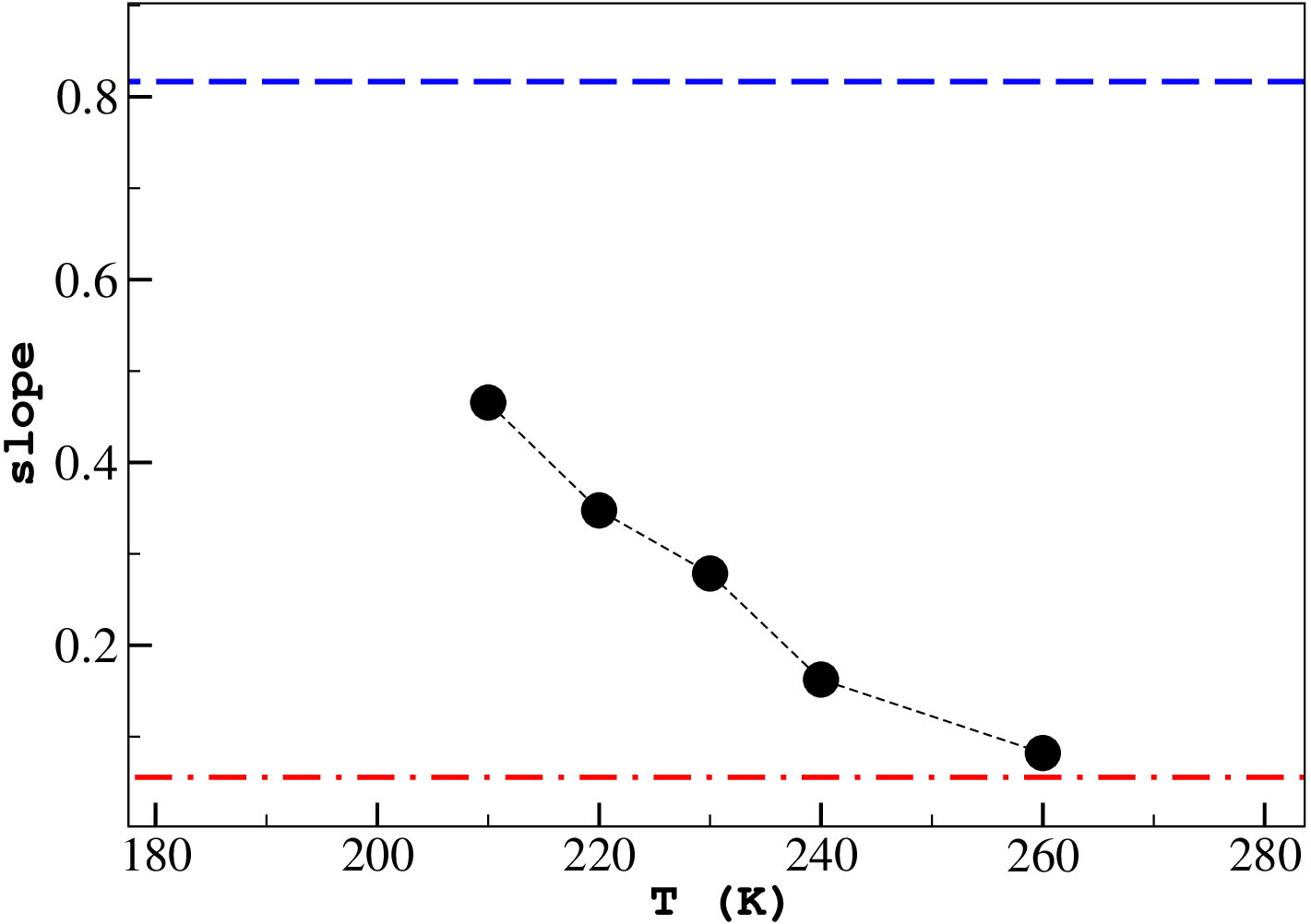}
\caption{
Slope of $\sigma_{_{MSD}}$ vs.  $MSD $  
as a function of $T$  for $\rho=$ 0.95 g/cm$^3$. 
The two lines indicate the two extreme limits provided by Eqs.~\ref{eq:highT}-\ref{eq:lowT}.}
\label{fig:slope}
\end{figure}

Fig.~\ref{fig:spread} shows
the behavior of the variance  $\sigma_{_{MSD}}$ 
as a function of the average $MSD$, 
parametrically in $t$, for several different $T$.
This representation, which accounts for the
intrinsic effect of the slowing down of the
dynamics on cooling by eliminating $t$, 
confirms that dynamic heterogeneities grow
significantly on supercooling.
Data in  Fig.~\ref{fig:spread} quantify that the spreading of the $MSD_i$ values is much more enhanced at low $T$. 
The $T$-dependence of the slope of the curves shown in Fig ~\ref{fig:spread} is shown in 
Fig.~\ref{fig:slope}.

It is interesting to compare results reported
in Fig.~\ref{fig:spread} and Fig.~\ref{fig:slope}
with expectations for a  
Gaussian random walk process. For this case,
the probability of finding the walker after time $t$ at distance $r^2$ from the 
location at the time origin is
\begin{equation}
P(r^2,t)dr^2 =   \sqrt{\frac{27} {2 \pi}} \frac{\sqrt{r^2}}
{<r^2(t)>^{3/2}} e^{-\frac{3 r^2}{2 <r^2(t)>}}dr^2.
\end{equation}
The first and second moment of this distribution are given by
\begin{equation}
<r^2>= \int_0^\infty r^2 P(r^2,t)dr^2=<r^2(t)>,
\end{equation}
and
\begin{equation}
<r^4>= \int_0^\infty r^4 P(r^2,t)dr^2 = \frac{5}{3} <r^2(t)>^2.
\end{equation}


The variance $\delta$  of $r^2$ between different trajectories of the walker is

\begin{equation}
\delta^2= <(r^2-<r^2>)^2> =\frac{2}{3}<r^2(t)>^2.
\end{equation}

Hence, a plot of $\delta$ vs. $<r^2(t)>$ has, for a single Gaussian walker, a slope of $\sqrt{\frac{2}{3}}$, an universal value independent on the diffusion constant.

 If the dynamics in simulated water could be represented by the dynamics of $N=216$ independent walker, then $\sigma_{MSD}$ should
 be related to $MSD$ by the relation
 \begin{equation}
\sigma_{MSD}= \sqrt{\frac{2}{3}} \frac{1}{\sqrt{N}} MSD.
 \label{eq:highT}
\end{equation}
\noindent
since  each $MSD_i$  would be the sum of
$N$ independent gaussian processes (with a reduction of the variance  by a factor  $\sqrt{N}$ as compared to the single random walker case).  Data in
Fig.~\ref{fig:spread} and Fig.~\ref{fig:slope}
shows that this limit is approached at high $T$.

In an other extreme theoretical case, each realization can be considered as a single random gaussian process  (for example, in the limit of strong correlation between all $N$ molecules, or in the limit of one single diffusing molecule).
In this limit, the expected relation 
between  $\sigma_{MSD}$ and $MSD$ would be
\begin{equation}
\sigma_{MSD}= \sqrt{\frac{2}{3}} MSD.
\label{eq:lowT}
\end{equation}
For reference, this limit is also reported in
Fig.~\ref{fig:spread} and Fig.~\ref{fig:slope}.

Data shown in Fig.~\ref{fig:spread} and Fig.~\ref{fig:slope} show that a cross-over from the behavior of Eq.~\ref{eq:highT} to the behavior of Eq.~\ref{eq:lowT} takes place on supercooling. It will be very interesting to study the size, $T$ and $t$ dependence of these effects. Moreover, since each of the equilibrium  trajectory of the 216 molecules system can be thought as representative part of a large system, the increase of the variance with decreasing $T$ provides  a strong evidence of growing dynamic  heterogeneities in  the system. In this respect, this set of data, or analogous data for simpler potentials, may become  a relevant tool for discriminating between different theories of the glass transitions,  in particular between the ones based  on facilitated dynamics ideas
\cite{chandler} and trap models\cite{dyre,bucheau}.

\section{Correlation between diffusion coefficient and $e_{IS}$}

\begin{figure}
\includegraphics*[width=\figwidth]{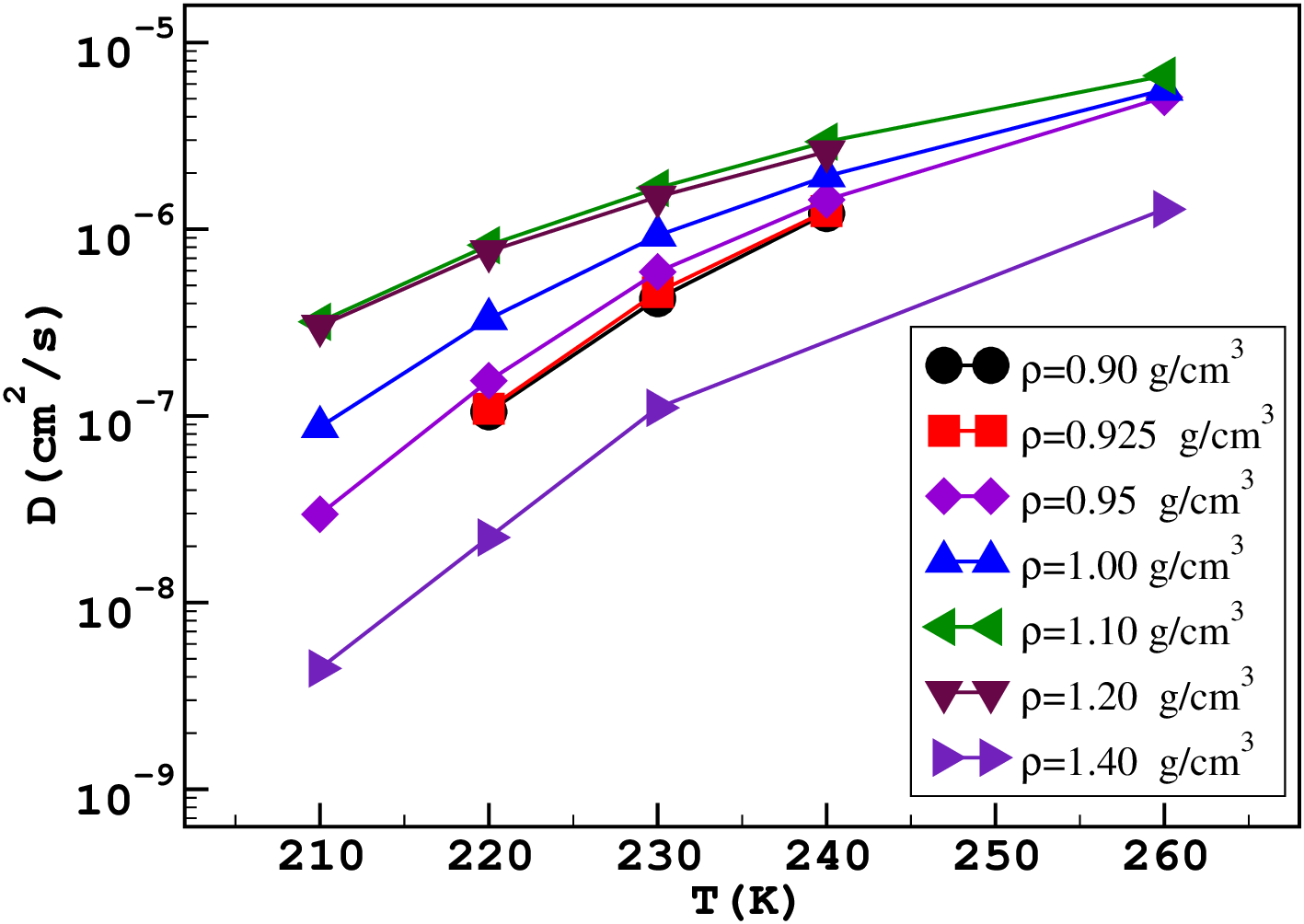}
\caption{Average diffusion coefficient $D$ as a  function of $T$ for several studied densities.}
\label{fig:diff}
\end{figure}

Bulk dynamic properties of SPC/E water have been previously 
 investigated in details\cite{gallo,spcemct,harrington,franciswater,pablo,fschemphys}.  
In particular,  PEL inspired studies have investigated the relation between
 the  $T$ and $\rho$ dependence of the diffusion coefficient $D$ and the number
 of unstable directions in configuration space\cite{inm}, as well as the relation
 between $D$ and the configurational entropy\cite{scala00}.  Fig. ~\ref{fig:diff} shows the average diffusion coefficient $D$, evaluated from the $MSD$ long time limit, for several studied density and temperatures. The present set of data improves the precision of previous estimate for the same model~\cite{franciswater}. Despite the relatively
small $T$-range investigated in this study, $D$ varies over more than three order of magnitudes.
It also clearly show that, among all the studied densities, $D$ is largest around density $1.1$ g/cm$^3$.  Increasing $\rho$,  dynamics slow down due to packing effect, while decreasing $\rho$  dynamics slow down due to the development  of a network of hydrogen bonds.

Most of previous studies on dynamic heterogeneities\cite{glotzer1,glotzer2,nicolas2} have focused on the  dynamics of different sub-set of molecules in the same system, raising sometime the 
question if the observed results were somehow associated with the chosen rules for identifying slow and fast particles.  It is also fairly difficult to correlate 
properties of the subset of molecules to energetic properties, due to the
difficulty of separating unambiguously the single particle energy contributions.
In the present approach, heterogeneities are addressed globally, as a fluctuation phenomenon.
To better clarify the connection between dynamic and energetic (static) heterogeneities, we  
 correlate the apparent diffusion coefficient of each trajectory with the corresponding average inherent structure.    Indeed, it is important to observe that, when each trajectory is averaged over different time origins,  the resulting $MSD$  behaves linearly with $t$, such that  an apparent trajectory diffusion coefficient $D_i$  can be estimated.  Differences in $D_i$ values between different trajectories persist even for time such that molecules have diffused over distances larger than two molecular diameters  (being $\approx 0.3$ nm the distance between the center of mass of two nearest neighbor molecules).

\begin{figure}[h]
\includegraphics*[width=\figwidth]{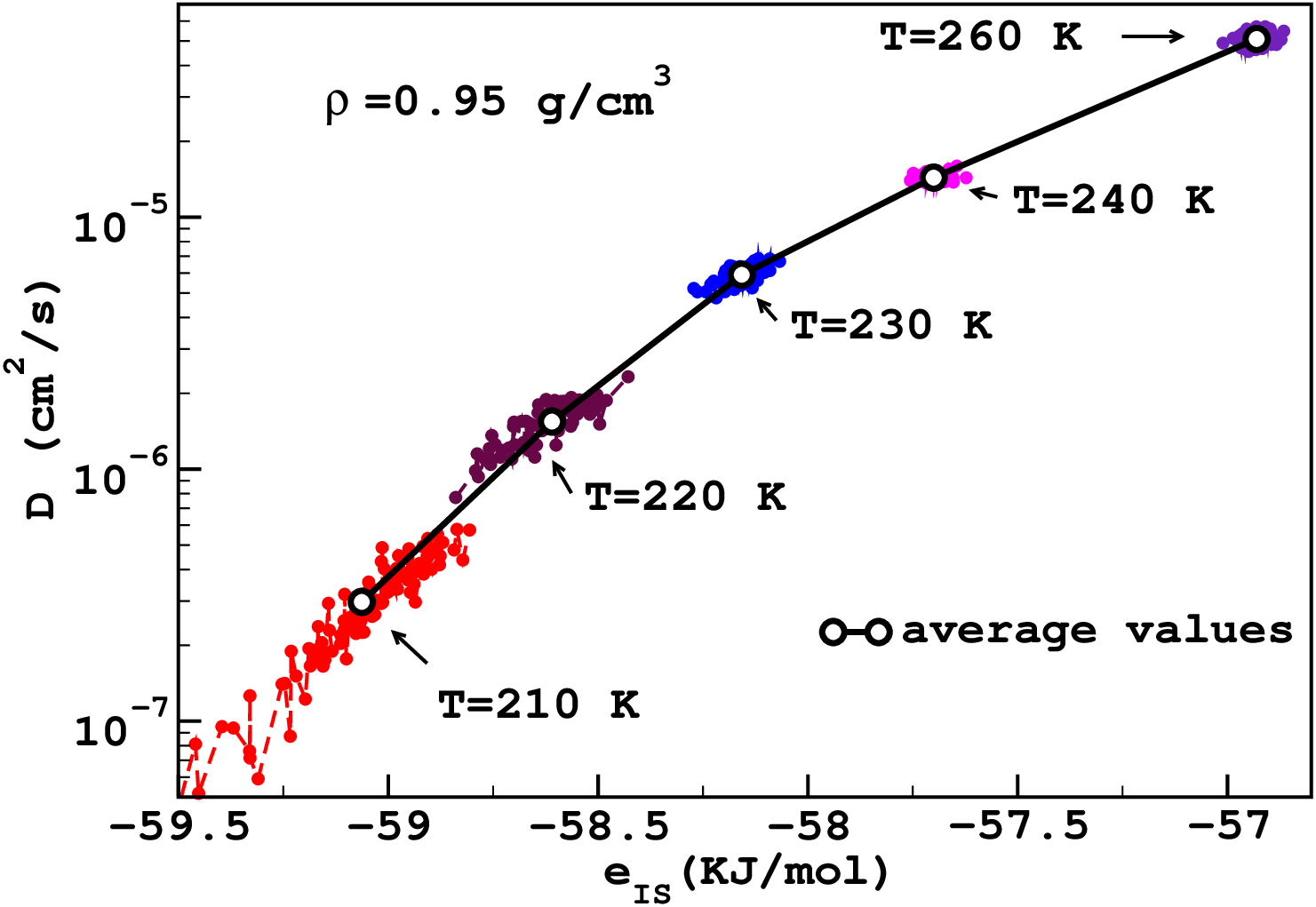}
\includegraphics*[width=\figwidth]{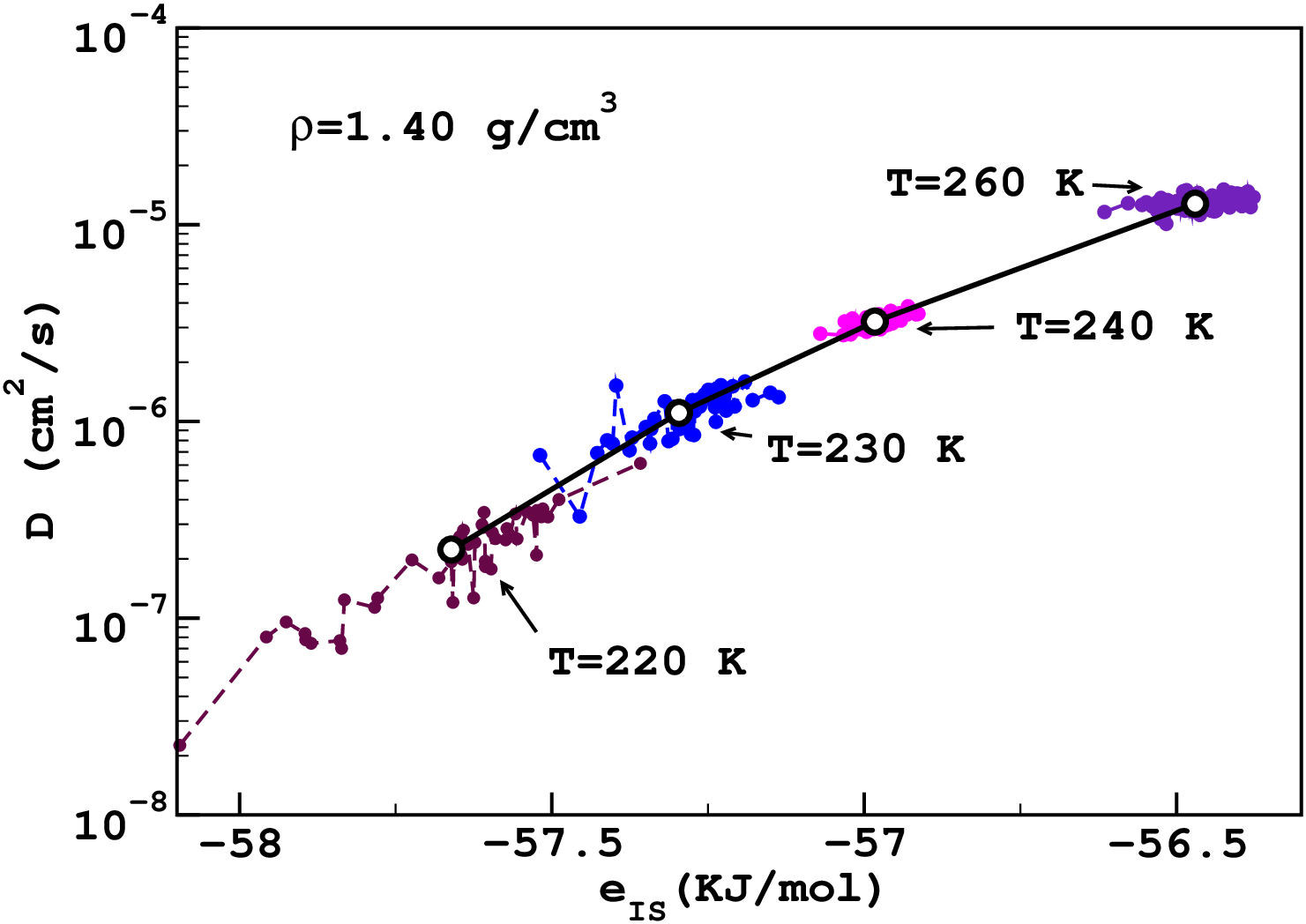}
\caption{Comparison between the average diffusion coefficient (line with open circles)  and the  diffusion coefficient  of each single trajectory at $\rho=0.95$ g/cm$^3$ (top panel) and $\rho=1.40$ g/cm$^3$ (lower panel). }
\label{fig:eisd095}
\end{figure}

 Fig.~\ref{fig:eisd095}  shows a  plot  of the average $D(T)$ as a  function  $<e_{IS}>(T)$ for one selected isochore. It also show for each trajectory $i$  ( at the same $T$ and $\rho$),  the apparent diffusion coefficient  $D_i$ vs the average sampled $<e_{IS}>_i$. While at larger $T$,  self-averaging is well accomplished on the time scale of the simulation,  at lower $T$, each trajectory ---  even if diffusive over distances of several molecular diameters --- samples only  a small part of the statistically relevant configurational space,  resulting in a large spreading in the value of $<e_{IS}>_i$.  Interestingly enough, such a spreading in $<e_{IS}>_i$ is strongly correlated to  the spreading in the  $D_i$ values.

The relation between  $D_i$ and  $<e_{IS}>_i$  displayed by the data is not very different from the corresponding relation $D$ vs $<e_{IS}>$ for the  averaged values.  A more detailed study of $D_i$ vs $<e_{IS}>_i$  over a smaller grid of temperatures, allowing for overlap of different $(D_i-<e_{IS}>_i)$ pairs of points could help sorting out the $T$ and  the $e_{IS}$ roles.   Indeed, one could associate each  $e_{IS}$ to a  precise value of   $D (e_{IS})$ and in analogy with the thermodynamics formalism developed by Stillinger,  one could attempt to calculate $D(T)$ as\cite{heuernew}
\begin{equation}
D(T) = \int D(e_{IS}) {\cal F}(e_{IS},T)  P(e_{IS},T) de_{IS}
\end{equation},
where ${\cal F}(e_{IS},T)$ account for the role of $T$ in dynamics 
and $P(e_{IS},T)$ is calculated according to Stillinger's PEL formalism. If this goal would be reached, the PEL formalism would become an exceptionally reach tool not only for describing the thermodynamics of supercooled liquids but also their dynamics.

\section{Conclusion}
\label{conclusions}

One of the key features in the slowing down of the dynamics of liquids on approaching 
supercooled states --- i.e. when dynamics begin to slow down significantly as compared to the standard  liquid values --- is the development of local fluctuations both in static and dynamic properties.  
The development of numerical resources allows us to start looking carefully into 
this problem, by performing analysis of the fluctuations\cite{ivan}.
These approaches, compared to  the corresponding studies of the average 
properties, are complicated by the interplay between space and time.
The data reported in this article, more than providing conclusive answers, 
hopefully clarify the richness of this type of analysis and stimulate further studies 
focusing on  the time and space evolution of the fluctuations.
 
The presented preliminary analysis reported clearly show that the role of fluctuations 
significantly grows on cooling.  Comparing fluctuations in dynamics  
at fixed mean square displacement (Fig.\ref{fig:spread}), we have detected a progressive increase of the fluctuations on cooling, already in the region of early supercooling, where MCT appears to provide a consistent description of the dynamics\cite{spcemct}. Interestingly, self-averaging properties appear to set in only after molecules 
have diffused several particle diameters.   This finding not only clarify the difficulty of calculating reliable values for  dynamical quantities in deep supercooled states but call attention on the fact that a  complete decorrelation of the system requires, at low $T$, rearrangements which  extends much beyond the first neighbor shell.

We have also observed a clear correlation between the apparent diffusion coefficient of the 
individual realizations and  depth of the sampled PEL (Fig.\ref{fig:eisd095}).
Again, the comparison of different trajectories, all at the same temperature, helps in eliminating trivial (but a priory unknown)  thermal effect, highlighting the connection between dynamics and IS energies.  
In this respect, one of the goal of future  studies should  be the
 development of a (size dependent) dynamical histogram reweighting formalism, 
conceptually similar  to the one used to calculate the density of states, from which 
the relation $D(e_{IS})$ could be extracted. 
This would allow us to sort out the role of $T$ and  $e_{IS}$  in dynamics,   
 and to describe in term of PEL properties not only the thermodynamics of supercooled liquids but also their dynamics.\\
\vskip 1cm
\begin{center}
{\bf ACKNOWLEDGEMENTS}
\end{center}
\noindent 
We thanks   MIUR-COFIN-2000 and FIRB for support. 
%
%

%

%
%

\begin{references}
%
\bibitem{wales}
D. J. Wales, {\it Energy Landscapes} (Cambridge University Press , Cambridge, 2003).

\bibitem{pablonature}
P. G. Debenedetti and F. H. Stillinger,
{ Nature } {\bf 410}, 259 (2001).

\bibitem{skt99}
F. Sciortino, W.  Kob and  P. Tartaglia, Phys. Rev Lett.  
{\bf 83}, 3214 (1999).

\bibitem{scala00}
A.~Scala, F.~W.~Starr, E. La Nave, F.~Sciortino and H.~E.~Stanley,
{ Nature} {\bf 406}, 166 (2000).
%
\bibitem{sastry01}
S.~Sastry, Nature  {\bf 409}, 164 (2001).
%
\bibitem{starr01}
F.~W.~Starr, S.~Sastry, E.~La Nave, A.~Scala, H.~E.~Stanley  and
F.~Sciortino, Phys. Rev. E {\bf 63}, 041201 (2001).
%
\bibitem{speedyjpc} R.~J.~Speedy, J. Chem. Phys. {\bf 114}, 9069 (2001). 
%
\bibitem{otplungo} S. Mossa, E. La Nave,
 H. E. Stanley, C. Donati, F. Sciortino  and
P. Tartaglia, Phys. Rev. E {\bf 65}, 041205-1 (2002).
%
\bibitem{wales2} D. J. Wales, Science {\bf 293}, 2067 (2001).
%
\bibitem{lacks}  D. J. Lacks, Phys. Rev. Lett. {\bf 80}, 5385 (1998).

\bibitem{heuernew}
B.  Doliwa  and A.  Heuer, { Phys. Rev. Lett.}
{\bf 91},  235501 (2003).
%
\bibitem{stillinger_pes}
F.~H.~Stillinger  and T.~A.~Weber, Phys. Rev. A {\bf 25}, 978 (1982);
Science {\bf 225}, 983 (1984); F.~H.~Stillinger, {\it ibid.} {\bf 267},
1935 (1995).
%
\bibitem{voivod01} I.~Saika-Voivod, P.~H.~Poole  and F.~Sciortino,
Nature (London) {\bf 412}, 514 (2001).

\bibitem{st}
F. Sciortino and P. Tartaglia,
{ Phys. Rev. Lett.} {\bf 86},  107 (2001).
\bibitem{crisanti}
A. Crisanti and  F. Ritort, { Europhys. Lett.} {\bf 52}, 640 (2000).

\bibitem{parrocchia}
S. Mossa, E. La Nave,  F. Sciortino and  P. Tartaglia,
{ Eurphys. J.  B} {\bf 30},  351  (2002).

\bibitem{scottjcp}
M. S. Shell, P. G. Debenedetti,   E. La Nave and F. Sciortino,
{ J. Chem.  Phys. }  {\bf 118},  8821  (2003).

\bibitem{spce}
H. J. C. Berendsen ,  J. R. Grigera and T. P. Stroatsma, 
{ J. Phys. Chem.} {\bf 91},  6269  (1987). 
 
\bibitem{t5p}
M.W. Mahoney and W. L. Jorgensen, J. Chem. Phys. {\bf 112},
8910 (2000); {\it ibid} {\bf 114},  363 (2001).

\bibitem{st2}
F. H. Stillinger and A. Rahaman, J. Chem. Phys. {\bf 60},  1545 (1974).

\bibitem{angellreview}
C. A. Angell,  in {\it Water: A Comprehensive Treatise} {\bf Vol. 7},
 (ed. Franks, F.) 1-81 (Plenum, New York, 1982).

\bibitem{pablobook}
P. G. Debenedetti, {\it Metastable liquids\/}.  (Princeton University
Press, Princeton, 1997).

\bibitem{poole}
P. H. Poole, F. Sciortino, U. Essmann and H. E. Stanley,
{ Nature} {\bf 360},  324 (1992).

\bibitem{ponyatonski} 
E.  G. Ponyatovsky,  V. V.   Sinitsyn  and  T. A. Pozdnyakova, 
{ JEPT Lett.} {\bf 60},  360  (1994).

\bibitem{lang}
F. X. Prielmeier, E. W. Lang, R. J. Speedy and H. D. Ludemann,
{ Phys. Rev. Lett.} {\bf 59}, 1128 (1987).

\bibitem{stillingerwater} F. H. Stillinger and T. A. Weber,
J. Phys. Chem. {\bf 87},  2833 (1983).

\bibitem{ohmine1}
I. Ohmine, H. Tanaka and  P. G. Wolynes,
{ J. Chem. Phys.} {\bf 89},  5852 (1988).

\bibitem{ohmine2}
M. Sasai, I. Ohmine and R. Ramaswamy,
{ J. Chem. Phys.} {\bf 96},  3045 (1992).

\bibitem{jcpmio}
F. Sciortino, A. Geiger and H. E. Stanley,
{ J. Chem. Phys.} {\bf 96}, 3857 (1992).


\bibitem{nature5neigh}
F. Sciortino, A. Geiger and H. E. Stanley,
{\it Nature} {\bf 354},  218 (1991).

\bibitem{debenewater}
C. J. Roberts, P. G. Debenedetti and  F. H. Stillinger 
{ J. Phys. Chem.  B} {\bf 103},  10258 (1999).


\bibitem{nicolas} 
N. Giovambattista , H. E. Stanley and  F.  Sciortino,
 { Phys. Rev. Lett.}, {\bf 91}, 115504 (2003).

\bibitem{genebirt}
E. La Nave , F. Sciortino, A. Scala, H. E. Stanley and F. W. Starr,
{ Europ. Phys. J E} {\bf 9},  233 (2002).

\bibitem{spceprl}
F. Sciortino, E. La Nave and P. Tartaglia,
 { Phys. Rev. Lett. }, {\bf 91}, 155701 (2003).

\bibitem{tom}
T. Keyes,
J. Chem. Phys. {\bf 101}, 5081 (1994). 

\bibitem{inm}
E. La Nave, A. Scala, F. Starr, H.E. Stanley and F. Sciortino,
{ Phys. Rev. Lett. } {\bf 84}  4605  (2000).

\bibitem{angelani}
L. Angelani, R. Di Leonardo, G. Ruocco, A. Scala, and F. Sciortino,
 { Phys. Rev. Lett. } {\bf 85},  5356 (2000).

\bibitem{silicaprl}
E. La Nave, H.E. Stanley and  F. Sciortino,
 { Phys. Rev. Lett. }  {\bf 88},  035501 (2002).

\bibitem{heuer}
A.~Heuer, Phys. Rev. Lett. {\bf 78}, 4051 (1997); S.~B\"uchner, and
A.~Heuer, Phys. Rev. E {\bf 60},  6507 (1999).
%
\bibitem{cavagna}
K. Broderix,  K.K. Bhattacharya, A.  Cavagna, A.  Zippelius  and  I. Giardina 
{ Phys. Rev. Lett. }  {\bf 85},  5360 (2000).

\bibitem{doliwa}
B. Doliwa B and A. Heuer, { Phys.  Rev. E} {\bf 67}, 030501 (2003).

\bibitem{schroder}
T.~B.~Schr{\o}der, S.~Sastry, J.~C.~Dyre and S.~C.~Glotzer, 
J. Chem. Phys. {\bf 112}, 9834 (2000).
%
\bibitem{harrowell}
H.~Fynewever, D.~Perera and  P.~Harrowell,
J. Phys. Condens. Mat. {\bf 12}, A399 (2000).
%

\bibitem{cristiano} C. De Michele, F. Sciortino and A. Coniglio, cond-mat 0405282 (2004).
%
\bibitem{ruoccojcp}
G. Ruocco,  F. Sciortino, F. Zamponi, C. De Michele and T. Scopigno,
 { J. Chem Phys} {\bf 120}, 10666 (2004).
%
\bibitem{pablo} 
C. J. Roberts, P. G. Debenedetti  and F. H. Stillinger,
{J.  Phys. Chem. B} {\bf 103 }, 10258 (1999).


\bibitem{numericalrec} Numerical recepies, 
\protect{$http://www.nr.com/nronline\_switcher.html$} .

\bibitem{sastry}  
 S. Sastry,  P. G.  Debenedetti,  F.   Sciortino and H. E. Stanley, 
{ Phys. Rev. E} {\bf 53}, 6144 (1996).


\bibitem{otppisa}
E. La Nave , F. Sciortino, P. Tartaglia, C. De Michele and S. Mossa,
J. Phys: Condens. Matter {\bf 15}, 1  (2003).


\bibitem{eos} 
E. La Nave, S.   Mossa   and  F. Sciortino,  
{ Phys. Rev. Lett.} {\bf 88},  225701[1] (2002).


\bibitem{sasai} 
M. Sasai,  
J. Chem. Phys. {\bf 118}, 10651 (2003).



\bibitem{rem} 
B.  Derrida,  
{ Phys. Rev. B} {\bf 24},  2613  (1981).

\bibitem{tomrem}
T. Keyes , J.  Chowdhary and J.  Kim, 
{ Phys. Rev. B} {\bf 66},  051110  (2002)

\bibitem{ivan}
I. Saika-Voivod   and  F. ~Sciortino, 
Phys. Rev. E, in press (2004).


\bibitem{tom2}
W. X. Li and T. Keyes,
{ J. Chem Phys} {\bf 111}, 328 (1999).

\bibitem{chandler}
J. P. Garrahan  and  D. Chandler,
Phys. Rev. Lett. {\bf 89},  035704  (2002).

\bibitem{dyre}
 J. C. Dyre   Phys. Rev. Lett. {\bf 58},  792 (1987);
 J. C.  Dyre  Phys. Rev.  B {\bf  51}, 12276 (1995).

\bibitem{bucheau}
                                                                                
C. M. and J.-P. Bouchaud ,  J. Phys. A: Math. Gen. {\bf 29},
3847-3869 (1996), and refs. therein.

\bibitem{gallo} 
F. Sciortino, P. Gallo, P. Tartaglia and S.H. Chen,
Phys. Rev. E {\bf 54}, 6331 (1996); P. Gallo,  F. Sciortino,  P. Tartaglia  and S. H.  Chen,
{ Phys. Rev. Lett.} {\bf 76},  2730  (1996).

\bibitem{spcemct}
L. Fabbian, A. Latz, R. Schilling, F. Sciortino, P. Tartaglia and C.  Theis,
{ Phys. Rev. E} {\bf 60}, 5768 (1999).

\bibitem{harrington}
S. Harrington , P. H. Poole , F. Sciortino  and H. E. Stanley,
{ J. Chem Phys} {\bf 107}, 7443 (1997).

\bibitem{franciswater}
F. W. Starr,  F. Sciortino and H. E. Stanley, 
{ Phys. Rev. E} {\bf 60},  6757 (1999).

\bibitem{fschemphys}
F. Sciortino, { Chem. Phys.} {\bf 258}, 307 (2000).


\bibitem{glotzer1}
C. Donati, J. F. Douglas, W. Kob, S. J. Plimpton, P. H. Poole  and S. C. Glotzer,
Phys. Rev. Lett. {\bf 80},  2338 (1998).

\bibitem{glotzer2}
W. Kob, C. Donati, S. J. Plimpton, P. H. Poole and S. C. Glotzer,
Phys. Rev. Lett. {\bf 79}, 2827-2830 (1997).

\bibitem{nicolas2}
N. Giovambattista, S. V.  Buldyrev, F. W.  Starr and  H. E.  Stanley,
Phys. Rev. Lett. {\bf 90},  085506  (2003).


\end{references}
\end{document}